\documentclass[aps,prc,floatfix,amsmath,12pt]{revtex4}
\usepackage{dcolumn}
\usepackage{amssymb}
\usepackage{mathrsfs}
\usepackage{textcomp}
\usepackage{bm}
\usepackage{supertabular}
\usepackage{color,graphicx}
\usepackage[bookmarksopen]{hyperref}
\newcommand{\bea}{\begin{eqnarray}}
\newcommand{\eea}{\end{eqnarray}}
\newcommand{\nn}{\nonumber\\}

\begin{document}
\title{Heavy quark damping rate in hot viscous QCD plasma}
\author{Sreemoyee Sarkar}
\email{sreemoyee.sarkar@saha.ac.in}

\author{Abhee K. Dutt-Mazumder}
\email{abhee.dm@saha.ac.in}

\affiliation{High Energy Nuclear and Particle Physics Division, Saha Institute of Nuclear Physics,
1/AF Bidhannagar, Kolkata-700 064, INDIA}

\medskip

\begin{abstract}
We derive an expression for the heavy quark damping rate in hot quark gluon plasma in presence of flow. 
Here all the bath particles are out of equilibrium due to the existence 
of non-zero velocity gradient. The magnetic sector shows similar infrared divergences even after hard thermal loop corrections as one encounters in case of 
non-viscous plasma. We estimate the first order correction in ($\eta/s$) for heavy quark damping rate due to the non-zero viscosity of the QCD plasma.

\end{abstract}

\maketitle
\section{Introduction}
It is now generally accepted that the matter produced in Relativistic Heavy Ion Collider experiments behave like near ideal fluid. 
In such collisions the ratio of the shear viscosity ($\eta$) and the entropy density ($s$) is estimated to be not more than 
few times the lower bound of $\eta/s=1/4\pi$ \cite{Kovtun05}. This conclusion 
has been drawn from the success of ideal hydrodynamics particularly in explaining both hadron transverse 
momentum spectra and elliptic flow ($v_2 (p_T)$) in the low $p_T$ region. 

The $v_2(p_T)$, is the second harmonic of the azimuthal 
distribution of the produced particles with respect to the reaction plane which is measured 
as a function of transverse momenta of various particle types and for different impact 
parameters. The ideal hydrodynamic description however breaks down
for $p_T> 2 GeV $ beyond which $v_2(p_T)$ does not rise as predicted by non-viscous hydrodynamics\cite{Teaney03,Dusling10}.

In ideal hydrodynamics $v_2(p_T)$ shows a rising trend with increasing $p_T$ \cite{Dusling10}. 
Recently several investigations have been performed to address this issue and it has been shown that the falling trend of $v_2(p_T)$ in the higher $p_T$  region can naturally be explained by invoking non-ideal (viscous) hydrodynamics. It is to be noted that in non-ideal fluid the energy momentum tensor 
($T^{\mu\nu}$) apart from the ideal (non-interacting) part will also receive a correction term $\delta T^{\mu\nu}$ involving both the coefficients of shear ($\eta$) and bulk viscosity ($\zeta$). Stating differently in 
non-ideal fluid, the sound attenuation length $\Gamma_s=(4\eta)/(3(\epsilon +p))$ is non-zero. However, the  attenuation length is small compared 
to the expansion rate $\tau$, where $\epsilon$ is the energy density and $p$ is the pressure of the 
fluid \cite{Teaney03}.

It is to be noted that a strong hydrodynamic response is possible when $\Gamma_s < \tau$. The non-zero $\eta$ (or $\Gamma_s$) modifies both the equation of motion and also the particle distribution functions. The latter now will have a viscous correction term $\delta f$ {\em i.e} for
the fluid particle distribution function and we shall write $f= f^0+\delta f$, where $f^0$ is the local
thermal distribution function \cite{Dusling10}. $\delta f$ involves the sound attenuation length which however can be determined 
by solving the linearized Boltzmann equation. In general $\delta f$ contains both the shear and the bulk viscosity 
coefficients, here, we restrict ourselves only to the modification of $\delta f$ due to $\eta$. 
A brief discussion on this will be presented in section IIA. 

One of the major activities in the area of high energy heavy ion experiments have been to understand how the dissipative effects modify various experimental observables like particle spectra, HBT radii or elliptic flow 
due to the modified energy momentum tensor or the viscous corrected distribution function \cite{Soff01,Adler01,Acdox02}. Recently the effect of non-zero $\eta$ on 
the photon and the dilepton production rates have been estimated \cite{Dusling08,Dusling10A,Bhatt10}. 
In case of photon the effect of non-zero viscosity leads to a larger thermalization time . 
According to the authors of \cite{Bhatt10} the role of non-ideal effects is to increase the net photon yield due to 
slowing down of the hydrodynamic expansion. For dilepton the space-time integrated transverse momentum spectra shows a
 hardening where the magnitude of the correction increases with the increasing invariant mass. 
Here the authors argue that the thermal description is reliable for an invariant mass $<(2\tau_0 T_0^2)/(\eta/s)$ \cite{Dusling08}, 
where, $\tau_0$ is the thermalization time and $T_0$ is the initial temperature. 
The effect of the viscous correction to the gluon dielectric function has also been studied \cite{Jiang10} with a
specific choice of the direction of the intermediate momentum exchange for the gluons. Attempt has
also been made to calculate the drag and diffusion co-efficient including the viscous corrections
numerically \cite{Das12}.

 In the present work we intend to calculate the heavy quark damping rate in presence of longitudinal flow with 
 viscous corrections upto ${\cal O}(\eta/s)$ which allows us to obtain closed form analytical results. The calculation of 
quark damping rate in equilibrium QED or QCD plasma has already been studied in past few years
 \cite{Pisarski92,Pisarski93,Blaizot96,Blaizot97,Peshier05}. 
From the study it has been established that in non-viscous medium the quark damping rate is plagued with
 infrared divergences. In the non-relativistic plasma, 
where one considers only the coulomb or electric interaction such divergences are removed by the Debye screening effects. The
problem becomes non-trivial in dealing with the relativistic plasma where one has to worry about both the electric and
the magnetic interactions. This additional complication actually arises due to the absence of the static magnetic screening. In
case of QED, in \cite{Blaizot96,Blaizot97} 
the authors have shown that the electric contribution to the quasiparticle
damping rate with plasma screening effects is finite and is of the order $\Gamma_{long}\sim g^2 T$ ($g$ is the 
coupling constant) whereas the transverse part remains 
divergent even after inclusion of the plasma corrections. This is because the latter is only dynamically screened. 
To obtain a finite result another resummation scheme has to be developed by using Bloch-Nordsieck propagator 
\cite{Blaizot96,Blaizot97}. In case of QCD one also encounters similar problem
in hot plasma where either one can use the magnetic mass for the gluons or adopt similar resummation scheme as 
developed in \cite{Blaizot96,Blaizot97}. The same problem was latter addressed in \cite{Boyanovsky98} by adopting renormalization group formalism. 
Both the formalisms however yield the same final result with non-perturbative corrections for the damping rate showing non-exponential 
time dependence.
It 
would be worthwhile to note that in degenerate plasma one obtains finite results with magnetic interactions without further
resummation unlike its high temperature counterpart. For detailed discussions 
about these issues we refer the readers to \cite{Manuel97, Manuel00, Sarkar10}. 

The viscous part is operative only when there exists momentum anisotropicity {\em i.e} we consider there exists a non-zero velocity gradient in such plasma. Naturally this
is a major departure from the above cited calculations where always the bath particles are assumed to be in equilibrium.
This is true only when there exists no velocity or temperature gradient in the plasma and there is no external force.

 \section{Formalism}
 
 In order to calculate the heavy quark damping rate in a viscous plasma, we first 
 recall what is done in case of non-viscous medium. There one starts with the Boltzmann kinetic equation given
 by
\bea
\left(\frac{\partial}{\partial t}+{\bf v_p}.\nabla_{\bf r} +{\bf F}.\nabla_{{\bf p}}\right)f_p=-\mathcal{C}[f_p],
\label{boltz}
\eea
here, the right hand side represents the collision integral which can be evaluated once the 
interactions are
known. ${\bf v_p}={\bf p}/  E_p$ is the particle velocity.

In absence of the external force and gradients of temperature, velocity or density Eq.(\ref{boltz}) takes very simple form,
\bea
\frac{\partial f_p}{\partial t}=-\mathcal{ C}[f_p].
\label{boltz_2}
\eea
The collision integral can be written as sum of multiple terms representing various scattering processes,
\bea
{\mathcal C[f_p]}=-({\mathcal C[f_p]_{1\rightarrow 2}}+{\mathcal C[f_p]_{2\rightarrow 2}}+{\mathcal C[f_p]_{2\rightarrow 3}}+\cdots).
\eea
 In the present work we are interested only in the $2\rightarrow 2$ 
processes and calculate only the damping rate leaving collisional energy loss calculation for future work \cite{AKDM13}. For 
two body interaction ($P+K\rightarrow P'+K'$), the explicit form of the collision 
integral is the following,
\bea
\mathcal{C}[f_p]&&=\frac{1}{2E_p}\int \frac{d^3k}{(2\pi)^3 2E_k}\frac{d^3p^{'}}{(2\pi)^3 2E_p'}\frac{d^3k^{'}}{(2\pi)^3 2E_k'}
\nn\
&\times& f_pf_k(1\pm f_{p^{'}})(1\pm f_{k^{'}})-f_{p^{'}}f_{k^{'}}(1\pm f_p)(1\pm f_k)\nn\
&\times&(2\pi)^4 \delta^4(P+K
-P^{'}-K^{'})
\frac{1}{2}\sum_{spin}{\cal| M |}^2.
\label{collision}
\eea
It is to be noted that in Eq.(\ref{collision}) all the distribution functions designated by $f_i$ for $i=k,k',p'$
 are either the Fermi or Bose distribution functions for the quarks or gluons respectively. The $\pm$ signs include both stimulated
  emission or the Pauli blocking respectively. 

In the present scenario we are concerned with the heavy 
  quark damping rate scattering off from quarks and gluons in the medium. The injected heavy quark now has 
both equilibrium and a fluctuating part ($f_p=f_p^0+\delta f_p$) and all the bath particles are in equilibrium. 
$f_i^0$ denotes the equilibrium fermion or boson distribution function. Inserting $f_p=f_p^0+\delta f_p$ in Eq.(\ref{boltz_2}) one can write,
\bea
\frac{\partial \delta f_p}{\partial t}&=&\frac{1}{2E_p}\int \frac{d^3k}{(2\pi)^32k }\frac{d^3p^{'}}{(2\pi)^32E_p' }\frac{d^3k^{'}}{(2\pi)^3 2k'}
f_{E_p}f_k^0(1\pm f^0_{k^{'}})\nn\
&\times&(2\pi)^4 \delta^4(P+K-P^{'}-K^{'})
\frac{1}{2}\sum_{spin}{\cal| M |}^2.
\label{collision_heavy_q}
\eea

Note the difference of the thermal phase space here with
that of the light quarks in \cite{AKDM05}. While writing the above equation for high energetic parton the possibility of 
back scattering has been excluded and the approximation $(1\pm f^0_{E_p'})\simeq 1$ has also been incorporated in the 
 thermal phase space since $E_{p'}>>T$. In the relaxation time approximation one can write,
 \bea
\frac{\partial \delta f_p}{\partial t}=-\mathcal{ C}[ f_p]=-\delta f_p \Gamma_p^0.
\label{boltz_3}
\eea

We can identify $\Gamma_p^0$ as the particle damping rate given by,
\bea
\Gamma_p^0&=&\frac{1}{2E_p}\int \frac{d^3k}{(2\pi)^3 2k}\frac{d^3p^{'}}{(2\pi)^3 2E_p'}\frac{d^3k^{'}}{(2\pi)^3 2k'}
 f_k^0(1\pm f_{k^{'}}^0) (2\pi)^4 \delta^4(P+K-P^{'}-K^{'})\frac{1}{2}\sum_{spin}{\cal| M |}^2.
\label{damp_non_vis}
\eea

 Note that $\Gamma_p^0$ is independent of non equilibrium part of 
$f_p$ and depends only on the distribution of the bath particles.

\subsection{Viscosity corrected distribution function}
To incorporate the effect of flow of the medium,
 we take viscous corrected distribution function as $f_i=f_i^0+\delta f_i^{\eta}$
 for the bath particles, where, $i=p^{'},k,k^{'}$. $\delta f_i^{\eta}$ is 
the first order correction to the thermal distribution function which actually 
is constrained by the viscosity or energy momentum stress tensor and also 
related to $\Gamma_s$. The form of $\delta f_i^{\eta}$ depends on the various ansatz \cite{Heiselberg93,
Heiselberg94,Arnold00,Dusling10}
\bea
\delta f_i^{\eta}=\chi(k)\frac{f_k^0(1\pm f_k^0)}{T}\hat k_i \hat k_j \nabla_i u_j.
\label{non_eq_dist_k}
\eea
In principle $\chi(k)$ can be determined from various microscopic theories as discussed in \cite{Dusling10}. 
In most of the hydrodynamic calculations it is assumed that $\delta f_k\propto k^2 f_k^0$ 
and the proportionality constant is independent of the particle type. 
This is known as quadratic ansatz which recently has been called into question \cite{Dusling10}. For QCD it has however been shown that
 $\chi(q)/\chi(g)\approx 1.70$. For the present case we assume these to be equal.

For a boost invariant expansion without transverse flow one can incorporate the viscous correction to 
the distribution function in the following way,
 \cite{Teaney03,Bhatt10,Das12},

 \bea
 \delta f^{\eta}_i(k)=f_i^0(1\pm f_i^0)\Phi_i(k),
 \label{vis_dist_func}
 \eea
 where,
\bea
\Phi_i(k)=\frac{1}{2T^3\tau}\frac{\eta}{s}\left(\frac{k^2}{3}-k_z^2\right).
\label{non_eqm_dist_func}
\eea
 The correction term given above is based upon the 
  "first approximation" described in \cite{Groot} and "one-parameter ansatz" for a variational solution of \cite{Arnold00}. The formal
  procedure for determining the viscous correction using variational principle has been discussed in \cite{Arnold00} 
in great detail. The viscous modification holds true only in the local rest frame 
 of the fluid and it contains the first order correction in the expansion of shear part
 of the stress tensor. $\tau$ is the thermalization time of the quark-gluon plasma (QGP) and the flow is along $z$ axis. From the
  above expression this is also evident that the non-equilibrium part of the distribution function 
  becomes operative only when there is a momentum anisotropy in the system. 
 
In a medium with non-zero flow gradient 
with the distribution functions of the form mentioned in Eqs.(\ref{vis_dist_func}) and (\ref{non_eqm_dist_func})
the collision integral can be expressed as,
\bea
\mathcal{ C}[f_p]&=&\frac{1}{2E_p}\int \frac{d^3k}{(2\pi)^32k }\frac{d^3p^{'}}{(2\pi)^32E_p' }
\frac{d^3k^{'}}{(2\pi)^3 2k'}\sum_{i=1,2}\alpha_i (2\pi)^4 
\delta^4(P+K-P^{'}-K^{'})\frac{1}{2}\sum_{spin}|{\mathcal M}|^2 ,
\eea
where, $\alpha_i$'s represent the viscous modified phase-space distribution functions. $\alpha_1$ 
 contains the equilibrium part of the distribution functions, this gives us the usual interaction rate mentioned in 
Eq.(\ref{damp_non_vis}) where all the bath particles are in thermal equilibrium,
\bea
\alpha_1=\delta f_pf_k^0(1\pm f_{k^{'}}^0).
\label{alpha_2}
\eea
$\alpha_2$ involves terms due to the viscous modifications to the light quark distribution functions for the bath constituents,
\bea
\alpha_2 &\simeq& \delta f_p\left[\Phi_kf_k^0(1\pm f_{k^{'}}^0)\pm \Phi_{k^{'}}f_{k^{'}}^0f_k^0\right].
\label{alpha_3}
\eea
In absence of viscosity $\Phi_i$'s are zero and we get back our result for 
usual quasiparticle damping rate which is given by Eq.(\ref{damp_non_vis}). In the relaxation time approximation
 using Eqs.(\ref{alpha_2}) and (\ref{alpha_3}) the collision
 integral can be written as,
 \bea
{\cal C}[f_p]\simeq\delta f_p\left(\Gamma^0(p)+\delta\Gamma^{\eta}(p)\right)
\label{rt_approx}
\eea
where $\delta\Gamma^{\eta}$ receives contribution from $\alpha_2$.

\subsection{Damping rate in presence of flow}
\begin{figure}[htb]
\centering
   \begin{tabular}
   {c@{\hspace*{12mm}}c@{\hspace*{8mm}}c@{\hspace*{8mm}}c}
\resizebox{3.5cm}{2.0cm}{\includegraphics{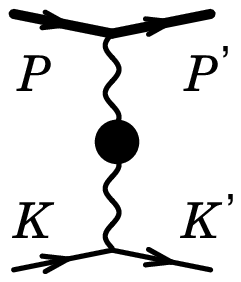}}

&
\resizebox{3.5cm}{2.0cm}{\includegraphics{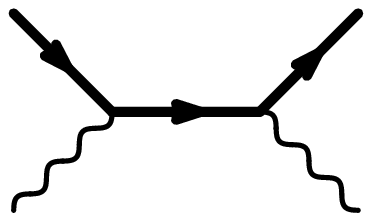}}
   &
    \resizebox{3.5cm}{2.0cm}{\includegraphics{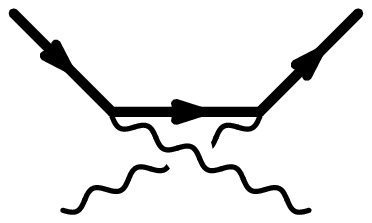}}
   &
\resizebox{3.5cm}{2.0cm}{\includegraphics{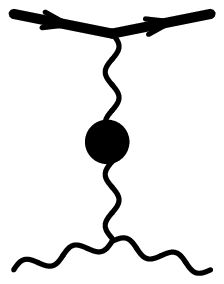}}
   \\
     (a) & (b) & (c) & (d)

  \end{tabular}
   \caption{Amplitudes for heavy quark elastic scattering in a QCD plasma. A curly line
 denotes a gluon (QCD). The amplitude (d) is specific to the QCD case.
 The blob in (a) and (d) denotes the resummed hard thermal loop boson propagator, which is
 necessary to screen the $t$-channel contribution in the infrared domain.}
   \label{fig:amp}
   \end{figure} 

It has already been indicated in the previous section that in presence of non-zero flow gradient particle 
distribution function gets modified with a term 
involving viscous corrections as given by Eqs. (\ref{non_eq_dist_k}), (\ref{vis_dist_func}) and (\ref{non_eqm_dist_func}). 
For the heavy quark or the test particle the distribution function also has a non-equilibrium 
fluctuating component $\delta f_p$ in addition to the equilibrium part $f_p^0$ ($\delta f_p<< f_p^0$).
It is to be noted that the heavy quark distribution function is independent of $\eta$.

In Eq.(\ref{collision}) we now insert the above mentioned viscous corrected distribution
 function to obtain,
\bea
\delta\Gamma^{\eta}_q(p)&=&\frac{1}{2E_p}\int_{k,p^{'},k^{'}} \left[\Phi_k f_k^0(1\pm f_{k^{'}}^0)\pm
\Phi_{k'}f^0_{k^{'}}f_{k}^0\right](2\pi)^4 
\delta^4(P+K-P^{'}-K^{'})\frac{1}{2}\sum_{spin}|{\cal M}| ^2,
\label{damp_vis}
\eea
 in the above equation $\int_k$ is shorthand for $\int d^3k/((2\pi)^32k)$. The above expression has been arrived at by neglecting terms ${\cal O}((\eta/s)^2)$ and ${\cal O}(f_i^3)$.
To proceed further we have to know the interaction. In case of quark-quark (q-q) scattering of different flavours in the $t$ channel and the quark-gluon (q-g) scattering in the same channel 
the matrix amplitudes squared
 are given by (see Fig.(\ref{fig:amp})) \cite{Arnold03},
 \bea
 \frac{1}{2}\sum_{spin}|{\cal M}|_{q_1q_2,t} ^2&=&\frac{8g^4d_F^2C_F^2}{C_A}\left(\frac{s^2+u^2}{t^2}\right),\nn\
 \frac{1}{2}\sum_{spin}|{\cal M}|_{qg,t} ^2&=&8g^4d_FC_FC_A\left(\frac{s^2+u^2}{t^2}+\frac{s^2+t^2}{u^2}\right).
 \label{mat_amp_bare_t}
 \eea
 The above matrix elements are singular because of the $t^{-2}=(\omega^2-q^2)^{-2}$ dependence, where $\omega$ and 
 $q$ are the energy
and momentum transfer. There are now two
singularities because of the Mandelstam variables, $u$ and $t$. For small momentum transfers the singular behavior can be cured with the help of the plasma screening
 which we discuss later in the present section. 

For the fermion exchange diagrams the matrix element is given by \cite{Arnold03},
\bea
 \frac{1}{2}\sum_{spin}|{\cal M}|_{qg,s+u} ^2&=&-8g^4d_FC_F^2\left(\frac{u}{s}+\frac{s}{u}\right).
\label{mat_amp_bare_su}
\eea
From the above matrix elements we take only the terms which give leading contributions to the heavy quark damping rate, hence, we 
approximately write, 
\bea
\frac{1}{2}\sum_{spin}|{\cal M}|_{t,qq} ^2&=&A_{b,q}\frac{\tilde s^2}{t^2}\nn\
\frac{1}{2}\sum_{spin}|{\cal M}|_{s+u,qg} ^2&=&-A_f\left(\frac{\tilde s}{\tilde u}+\frac{\tilde u}{\tilde s}\right),
\label{mat_amp_bare_stu}
\eea

for heavy quark $\tilde s= s-m_q^2$, $\tilde u=u-m_q^2$. The group factor for $t$ channel diagrams are given by 
$A_{b,q}=\frac{d_F^2C_F^2}{C_A}$ and 
$A_{b,g}=d_FC_FC_A$, where, $A_{b,q}$ is relevant for quark-quark and $A_{b,g}$ for quark-gluon scattering. $A_f$ for $s$ and $u$ channel diagrams is $A_f=d_FC_F^2$.
The divergences in the $t$ channel diagrams are usually removed with the help of the plasma screening.
 The usual way to include this medium modification is to use the Hard Thermal Loop (HTL) 
 dressed propagator instead of bare one \cite{Pisarski89,Pisarski90,Braaten91}. 
 With the dressed gluon propagator the quark-quark matrix amplitude squared looks like,
 \bea
\frac{1}{2}\sum_{spin}|M|_{qq}^2=8A_{b,q}g^4\left[\frac{1}{\left(q^2+\Pi_L\right)}-
\frac{v_{p,T}.v_{k,T}}{(q^2-\omega^2+\Pi_T)}\right]^2.
 \label{mat_amp_dressed}
 \eea
 In the above equation the medium modified gluon propagator
contains the polarization functions $\Pi_L$
 and $\Pi_T$, which describe plasma screening of interparticle
interaction by longitudinal and transverse plasma perturbations,
respectively. In the large wavelength limit $q<<T$ \cite{Heiselberg93},
\bea
\Pi_L (q,\omega)=m_D^2\chi_L,\quad \Pi_T (q,\omega)=m_D^2\chi_T,
\label{pol_tensor}
\eea
where,
\bea
\chi_L&=&\left[1-\frac{x}{2}\mbox{ln}\left(\frac{x+1}{x-1}\right) \right],\nn\
\chi_T&=&\left[\frac{x^2}{2}+\frac{x(1-x^2)}{4}\mbox{ln}\left(\frac{x+1}{x-1}\right) \right],
\label{pol_functions}
\eea
and $m_D$ is the gluon Debye mass. In the static limit ($x=\omega/q\sim 0$) above mentioned polarization functions can be expanded to give
 rise to $\chi_L\sim 1$ and $\chi_T\sim \frac{i\pi x}{4}$ when $|x|<<1$. In this region the squared
  matrix element becomes,
 \bea
\frac{1}{2}\sum_{spin}| M|^2_{qq} =8A_{b,q}g^4\left[\frac{1}{\left(q^2+m_D\right)^2}+
\frac{\left(v_p^2-x^2\right)q^2\mbox{cos}^2\phi}{\left(1-x^2\right)\left(q^6+\frac{\pi^2\omega^2m_D^4}{16}\right)}\right],
\label{mat_amp_cross_term}
\eea

 where, we have omitted the cross term of longitudinal and transverse interaction because azimuthal angle $\phi$ integration gives 
 zero contribution for this term. With the above matrix amplitude squared we now compute viscous corrected quark damping rate for the quark-quark scattering
 $\delta \Gamma_{qq}^{\eta}$,
\bea
\delta\Gamma_{qq}^{\eta}&\simeq&\frac{1}{2E_p}\int_{k,p,p'}\Big[\Phi_kf_k^0(1-2f_k^0+2\omega f_k'^0)+\omega \Phi_k'f_k^{02}+{\cal O}(\omega^2)\Big]\nn\
&\times&\frac{1}{2}\sum_{spin}|{\mathcal M}|^2
(2\pi)^4 
\delta^4(P+K-P^{'}-{K^{'}})
\label{all_terms}
\eea

while writing the above equation we have used the following expansion,
\bea
f^0(k')&=&f^0(k-\omega)\simeq f_k^0-\omega f_k'^0,\nn\
\Phi_{k'}&=&\Phi_{(k-\omega)}\simeq \Phi_k-\omega \Phi_k'.
\label{expansion}
\eea
To simplify $\delta\Gamma_{qq}^{\eta}$ we neglect higher order terms in $\omega$. With the bare interaction one observes that the 
above mentioned damping rate has infrared divergences and also the order of divergences for different terms are different.
 Both $\int dq/q^3$ and $\int dq/q$ divergences are present. As mentioned
   earlier in this section the usual way to handle these divergences is to incorporate the effects of plasma
    screening. The method of calculating the effects of screening
developed by Braaten and Yuan \cite{Yuan91} involves introducing
an arbitrary momentum scale $q^*$ to separate the region of
hard momentum transfer $q\sim T$ from the soft region
$q\sim gT$. The arbitrary momentum scale is chosen in the way so that $gT<<q^*<<T$, which
is possible in the weak-coupling limit $g\rightarrow0$.  The contribution from the hard momentum region $q>q^*$ is calculated using
tree-level scattering diagrams where the lower limit $q^*$ acts as infrared cutoff.
 
The detailed calculations of all the terms in Eq.(\ref{damp_vis}) have been presented in the 
Appendix A and B. Here, we only quote the final results. The final expressions for both the electric and the magnetic
 sectors in case of q-q scattering in the $t$ channel take the following forms,
 \bea
\delta\Gamma_{qq}^{\eta,L}&=&\left(\frac{\eta}{s}\right)\frac{g^4 A_{b,q}}{4(2\pi)^3\tau v_p}\left[\pi^2\left(
-\frac{ v_p}{6}-\frac{v_p^3}{9}+\frac{v_p^5}{10}\right)+\zeta(3)\left(\frac{3}{4} v_p +\frac{3}{2}
v_p^3 -\frac{21}{20} v_p^5 \right)\right]\nn\
&\times&\left(-\frac{1}{2}+\ln\left(\frac{q_{max}}{m_D}\right)\right),\nn\
\delta\Gamma_{qq}^{\eta,T}&=&\left(\frac{\eta}{s}\right)\frac{ A_{b,q}g^4\pi T^2}{16(2\pi)^3\tau v_pm_D^2}\left(\frac{7 \pi ^4}{15}-\frac{45 \zeta (5)}{2}\right)
\int_0^{q^*} \frac{dq}{q}\nn\
&+&\left(\frac{\eta}{s}\right)\frac{ A_{b,q}g^4}{8(2\pi)^3\tau v_p}
\Bigg[\left(\frac{\pi ^2}{3}-3 \zeta (3)\right)
\left(-\frac{16v_p^5}{225}-\frac{4v_p^5}{15}\ln\left|\frac{2q_{max}}{m_D\sqrt{\pi v}}\right|\right)\Bigg].
\label{final_result_qq}
\eea
In the high energy limit $q_{max}\sim \sqrt{E_pT}$ \cite{Beraudo06}.
From the above equations it is evident that the transverse sector contains both the finite and the 
infrared divergent terms. It is to be noted that presence of $\omega$ in the numerator in Eq.(\ref{all_terms})
 makes some of the divergent terms finite once the integrations are performed. Still however $\int dq/q$
 divergence remains for the first two terms in Eq.(\ref{all_terms}). This is reminiscent of what happens
  for the case of quasiparticle damping rate in non-viscous plasma which requires further resummation as 
  discussed in \cite{Blaizot96,Blaizot97}.

The physical processes
responsible for these divergences are the collisions involving the exchange of long wavelength, quasistatic,
magnetic gluons, which are not screened by plasma effects and show logarithmic divergence.
The leading divergences can be resummed
using a nonperturbative treatment based on a generalization of the Bloch-Nordsieck model at 
finite temperature \cite{Blaizot96,Blaizot97}. The resulting expression of the fermion propagator is free of infrared 
divergences, and exhibits a nonexponential
damping at large times \cite{Blaizot96,Blaizot97}.

Later Boyanovsky {\em et al} have shown that result obtaind 
 in \cite{Blaizot96,Blaizot97} 
 can be reproduced by invoking the renormalization group method \cite{Boyanovsky98}. This
allows a consistent resummation of infrared effects associated with the exchange of quasistatic transverse
gluons leading to anomalous logarithmic relaxation of the form $e^{-\alpha T t \ln[\omega_p t]}$ for hard momentum 
excitations where $\alpha$ is the fine structure constant and $\omega_p$ is the plasma frequency. 

In this context we recall the result of the heavy quark damping rate in non-viscous medium. It has been calculated long ago by Thoma
 and Gyulassy \cite{Thoma91} in hot QCD plasma. Here, we quote the final result,
 \bea
 \Gamma(p)_{qq}^{0,L}=\frac{g^4A_{b}T^3}{96\pi }\left(\frac{1}{m_D^2}-\frac{1}{q_{max}^2}\right)\nn\
 \Gamma(p)_{qq}^{0,T}=\frac{g^4A_{b}T^3 v_p}{48\pi m_D^2 }\int_0^{q^*}\frac{dq}{q}. 
 \label{non_vis_damp}
 \eea
 Thus we see that both the viscous corrected and non-viscous distribution functions for the bath particles
 give similar divergent result without any additional divergence. 

We now compute the contribution ($\delta\Gamma_{qg}^{\eta}$) to the heavy quark damping rate from Compton scattering.
 We start with Eq.(\ref{damp_vis}), where, the matrix element for q-g scattering including all three channels
  ${\cal M}={\cal M}_s+{\cal M}_t+{\cal M}_u$ has to be inserted. First we estimate the contributions from the $s$ and $u$
   channel diagrams,
   \bea
\delta\Gamma^{\eta}_{qg}&=&\left(\frac{\eta}{s}\right)\frac{A_fg^4}{32T^3\tau \pi^3 E_p}
\left[\frac{2\pi^4T^4}{135}-\frac{T^3m_q^2}{3E_p}\left
(2\zeta(3)\ln\left|\frac{4E_pT}{m_q^2}\right|+3\zeta(3)-2\gamma\zeta(3)+2\zeta'(3)\right)\right].
 \label{final_result_qg}
 \eea
 The contribution of quark-gluon scattering in heavy quark damping rate in a non-viscous medium has the following form,
 \bea
 \Gamma(p)^0_{qg}&=&\frac{1}{16\pi E_p}\int \frac{d^3k f_k}{(2\pi)^32k}\int dt\frac{{\cal|M|}^2}{s-m_q^2}\nn\
 &=&\frac{g^4T^2A_f}{48\pi E_p}\left[\ln\left|\frac{4E_pT}{m_q^2}\right|+{\cal O}(1)\right].
 \eea
 The final expression for the heavy quark damping rate can be obtained by adding the contributions from q-q 
 (Eq.(\ref{final_result_qq}) and q-g (Eq.(\ref{final_result_qg})) scatterings,
 \bea
 \delta\Gamma^{\eta}_q&=&\delta\Gamma_{qq}^{\eta}+\delta\Gamma_{qg}^{\eta}\nn\
&=&\left(\frac{\eta}{s}\right)\frac{g^4 A_{b}}{4(2\pi)^3\tau v_p}\left[\pi^2\left(
-\frac{ v_p}{6}-\frac{v_p^3}{9}+\frac{v_p^5}{10}\right)+\zeta(3)\left(\frac{3}{4} v_p +\frac{3}{2}
v_p^3 -\frac{21}{20} v_p^5 \right)\right]
\left(-\frac{1}{2}+\ln\left(\frac{q_{max}}{m_D}\right)\right)\nn\
&+&\left(\frac{\eta}{s}\right)\frac{ A_{b}g^4\pi T^2}{16(2\pi)^3\tau v_pm_D^2}\left(\frac{7 \pi ^4}{15}-
\frac{45 \zeta (5)}{2}\right)
\int_0^{q^*} \frac{dq}{q}\nn\
&+&\left(\frac{\eta}{s}\right)\frac{ A_{b}g^4}{8(2\pi)^3\tau v_p}
\Bigg[\left(\frac{\pi ^2}{3}-3 \zeta (3)\right)
\left(-\frac{16v_p^5}{225}-\frac{4v_p^5}{15}\ln\left|\frac{2q_{max}}{m_D\sqrt{\pi v_p}}\right|\right)\Bigg]\nn\
&+&\left(\frac{\eta}{s}\right)\frac{A_fg^4}{32T^3\tau \pi^3 E_p}\left[\frac{2\pi^4T^4}{135}-
\frac{T^3m_q^2}{3E_p}\left
(2\zeta(3)\ln\left|\frac{4E_pT}{m_q^2}\right|+3\zeta(3)-2\gamma\zeta(3)+2\zeta'(3)\right)\right].
 \eea
 where, we have defined $A_{b}=A_{b,q}+A_{b,g}$.

 An interesting implication of the characteristic behaviour of the damping rate as obtained above is closely related to the
 heavy ion experiments where non-equilibrated heavy quark passes through the longitudinally expanding QGP medium. 
Damping rate eventually determines how rapidly non-equilibrated heavy quark approaches its equilibrium state. 
 \section{Summary and Conclusion}  
 In the present work we have calculated heavy quark damping rate in a viscous medium restricting ourselves only to two body
  scattering processes {\em i.e} we consider only the quark-quark and quark-gluon scatterings. It has been shown in the text,
 how does the viscosity enter into the calculation {\em via} the viscous corrected distribution function 
 in the phase-space factor in presence of a flow gradient.
  We have restricted ourselves only to the leading order contributions in $\eta/s$ by dropping all the 
  higher order terms. To further simplify the collision integral the powers beyond the quadratic terms of the distribution functions have been dropped. Furthermore,
   for the gluon exchange only the soft frequencies have been retained following the standard techniques 
   what one adopts to calculate the damping rate
    formalism in non-viscous plasma. 
These approximations, in effect, allow us to present closed form analytical results. The final result is based upon the ansatz 
mentioned in Eqs.(\ref{vis_dist_func}) and (\ref{non_eqm_dist_func}). It is to
     be noted for the flow we consider only the longitudinal gradient by assuming that there is no transverse 
expansion. One of the
      interesting findings of the present work has been the infrared behaviour of the 
transverse damping rate which has the same 
      $\int dq/q$ form both for the viscous and the non-viscous part. To cure this divergence, which remains even after using finite
       temperature HTL propagator due to the non-existence of screening for the static gluons one may 
       perform further resummation by using Bloch-Nordsieck propagator or renormalization group method 
       \cite{Blaizot96,Blaizot97,Boyanovsky98} as mentioned in the text.

\appendix*{}
\section{Damping rate calculation}
\subsection{Quark-quark scattering}
In this section we explicitly show the computation of the q-q and q-g scattering rates in the $t$ channel 
in a viscous medium. First we evaluate the $t$ channel diagram for q-q scatterings. For this we start from the
 expression Eq.(\ref{damp_vis}),
\bea
\delta\Gamma_{qq}^{\eta}&\simeq&\frac{1}{2E_p}\int_{k,p,p'}\Big[\Phi_kf_k^0(1-2f_k^0+2\omega f_k'^0)
+\omega \Phi_k'f_k^{02}+{\cal O}(\omega^2)\Big]\nn\
&\times&\frac{1}{2}\sum_{spin}|{\mathcal M}|^2
(2\pi)^4 
\delta^4(P+K-{P^{'}}-K^{'})\nn\
&=&\delta\Gamma_1+\delta\Gamma_2+\delta\Gamma_3+\delta\Gamma_4.
\label{four_terms}
\eea
Considering the first two terms of the coefficient $\Phi_k$,

\bea
\delta\Gamma_{1+2}(p)&=&\left(\frac{\eta}{s}\right)\frac{1}{2^4(2\pi)^5×2T^3\tau}\int \frac{d^3p^{'}d^3k}
{E_pE_{p'}kk'}
f_k^0\left(1-2f_k^0\right)\left(\frac{k^2}{3}-k_z^2\right)\nn\
&&\delta(E_p+k-E_{p'}-k')
\frac{1}{2}\sum_{spin}|{\mathcal M}|^2.
\eea
In the collision integral we use the spatial delta function to perform the ${\bf k^{'}}$ integration and to shift the ${\bf p^{'}}$ integration
 into an integration over ${\bf q}$, where, ${\bf q}={\bf p^{'}}-{\bf p}$. 
It is convenient to introduce a dummy integration variable $\omega$,
\bea
\delta(E_p+k-E_{p'}-k^{'})=\int_{-\infty}^{\infty}d\omega 
\delta(\omega+E_p-E_{p'})\delta(\omega-k+k^{'}).
\eea
Evaluating ${\bf q}={\bf p^{'}}-{\bf p}$ in terms of p, q and $\mbox{cos} \theta_{pq}$ and defining $t=\omega^2-q^2$ we find,
\bea
\delta(\omega+E_p-E_{p'})&&=\frac{1}{v_pq}\delta\left(\mbox{cos} \theta_{pq}-\frac{\omega}{v_pq}-\frac{t}{2pq×}\right),\nn\
\delta(\omega-k+k^{'})&&=\frac{1}{q}\delta\left(\mbox{cos} \theta_{kq}-\frac{\omega}{q}+\frac{t}{2kq×}\right).
\eea
Using the above delta functions one can arrive at the following equation,
\bea
\delta\Gamma_{1+2}(p)&=&\left(\frac{\eta}{s}\right)\frac{1}{(2\pi)^42^42T^3\tau v_p}\int 
\frac{dqd\omega d\mbox{cos}\theta_{q}k^2dk d\mbox{cos}
\theta_kd\phi_k}{E_pE_{\bf{p+q}}k{\bf(k-q)}}f_k^0\left(1-2f_k^0\right)\left(\frac{k^2}{3}-k_z^2\right)\nn\
&\times&\delta\left(\mbox{cos} \theta_{pq}-\frac{\omega}{v_pq}-
\frac{t}{2pq×}\right)\delta\left(\mbox{cos} \theta_{kq}-\frac{\omega}{q}+\frac{t}{2kq×}\right)
\frac{1}{2}\sum_{spin}|{\mathcal M}|^2 .
\eea

To perform the phase space integration we choose $q$ along the $z$ axis, $p$ in the $x-z$ plane
and $k$ remains arbitrary,
\bea
	\bm q &=& (0,0,1)q \, , \nonumber \\
	\bm p &=& (\sin\theta_p,0,\cos\theta_p)p \, , \nonumber \\
	\bm k &=& (\sin\theta_{k}\cos\phi_{k},\sin\theta_{k}\sin\phi_{k},\cos\theta_{k})k \, .
\label{eq:angles_a}
\eea

Using the first delta function integration over $d\mbox{cos}\theta_{q}$ can be done.
 The delta function imposes the following condition on the angle $\theta_{pq}$,
\bea
\hat p. \hat q=\mbox{cos}\theta_{pq}=\frac{\omega}{v_pq}+\frac{t}{2pq}.
\label{delta_p}
\eea 
Second delta function yields,
\bea
\hat k. \hat q=\mbox{cos}\theta_{kq}=\frac{\omega}{q}-\frac{t}{2kq}.
\label{delta_k}
\eea
With the help of the first delta functions the bounds on $\omega$ can be fixed as follows,
\bea
\omega_{\pm}=E_p-\sqrt{E_p^2+q^2\mp 2E_pv_pq}.
\label{bounds_omega}
\eea
In case of a high energetic quark {\em i.e} when $E_p>>q$, the above limits can be approximated as $\omega\approx\pm v_pq$.
Using the second delta function we explicitly write the term coming from the viscous corrected distribution function as follows,
\bea
\left(\frac{1}{3}-\cos^2\theta_{kq}\right)=\left(\frac{1}{3}-\chi\right).
\eea

First we consider the soft sector of Eq.(\ref{four_terms}),
 \bea
\delta\Gamma_{1+2+3+4}^{soft}&=&\left(\frac{\eta}{s}\right)\frac{A_{b,q}g^4}{(2\pi)^34T^3\tau v_p}
\int_0^{\infty}k^4dk
\int_0^{q^*} dq\int_{-v_pq}^{v_pq}d
\omega \Bigg[\left(f_k^0-2f_k^{02}+2\omega f_k^0f_{k'}^0\right)\left(\frac{1}{3}-\chi\right)\nn\
&+&\Phi_k'\omega f_k^{02}\Bigg]
\left[\frac{1}{(q^2+m_D^2)^2}+\frac{\left(v_p^2-\frac{\omega^2}{q^2}\right)}{2\left(1-\frac{\omega^2}{q^2}\right)}\frac{1}{
\left(q^4+\frac{\pi ^2m^4\omega^2}{4q^2}\right)}\right].
\label{long_trans_tot}
\eea
We use following results for the $\omega$ integration to proceed further,
\bea
&&\int_{-v_pq}^{v_pq} \left(\frac{1}{3}-\left(\frac{\omega }{q}-\frac{\omega^2-q^2}{2kq}\right)^2\right) \, d\omega \nn\
&&=\frac{q^3 }{ k^2}\left(\frac{-v_p}{2}+\frac{v_p^3}{3}-\frac{v_p^5}{10}\right)+\frac{2qv_p}{3}
\left(1-v_p^2\right),\nn\
&&\int_{-v_pq}^{v_pq} \left(\frac{\omega}{3}-\omega\left(\frac{\omega }{q}-\frac{\omega^2-q^2}{2kq}\right)^2\right) \, d\omega\nn\
&&=\frac{q^3}{ k}\left(-\frac{2v_p^3}{3}+\frac{2  v_p^5}{5}\right).
\label{identity}
\eea
Hence, for the first two terms in Eq.(\ref{long_trans_tot}) in the electric sector we get,
\bea
\delta\Gamma_{1+2,l}^{soft}&=&\left(\frac{\eta}{s}\right)\frac{ A_{b,q}g^4}{(2\pi)^34T^3\tau v_p}
\Bigg[\left(\frac{-v_p}{2}+\frac{v_p^3}{3}-\frac{v_p^5}{10}\right)
\int_0^{\infty}k^2(f_k^0-2f_k^{02})dk
\int_0^{q^*}\frac{q^3dq}{(q^2+m_D^2)^2}\nn\
&+&\left(\frac{-2v_p^3}{3}+\frac{2v_p^5}{5}\right)
\int_0^{\infty}k^4(f_k^0-2f_k^{02})dk
\int_0^{q^*}\frac{qdq}{(q^2+m_D^2)^2}\Bigg]\nn\
 &=&\left(\frac{\eta}{s}\right)\Bigg[\frac{A_{b,q}g^4}{(2\pi)^34\tau v_p}\left(\frac{-v_p}{2}+\frac{v_p^3}{3}
-\frac{v_p^5}{10}\right)\left(
\frac{\pi^2}{3}-\frac{3\zeta(3)}{2}\right)\left(-\frac{1}{2}-\ln\left|\frac{m_D}{q^*}\right|
\right)\nn\
&+&\frac{A_{b,q}g^4T^2}{(2\pi)^34\tau v_p}\left(\frac{-v_p^3}{3}+\frac{v_p^5}{5}\right)
\left(\frac{7\pi^4}{15}-\frac{45\zeta(5)}{2}\right)
\left(\frac{1}{m_D^2}-\frac{1}{q^{*2}}\right)\Bigg].
\eea
Hard gluon exchange gives,
\bea
\delta\Gamma_{1+2,l}^{hard}&=&\left(\frac{\eta}{s}\right)\Bigg[\frac{A_{b,q}g^4}{(2\pi)^34\tau v_p}
\left(\frac{-v_p}{2}+\frac{v_p^3}{3}-\frac{v_p^5}{10}\right)\left(
\frac{\pi^2}{3}-\frac{3\zeta(3)}{2}\right)\left(\ln\left|\frac{q_{max}}{q^*}\right|
\right)\nn\
&+&\frac{A_{b,q}g^4T^2}{(2\pi)^34\tau v_p}\left(\frac{-v_p^3}{3}+\frac{v_p^5}{5}\right)\left(\frac{7\pi^4}{15}-\frac{45\zeta(5)}{2}\right)
\left(\frac{1}{q^{*2}}-\frac{1}{q_{max}^2}\right)\Bigg].
\eea

In the high energy limit from the kinematics $q_{max}$ can be taken as $\sqrt{E_pT}$ \cite{Beraudo06}, hence,
\bea
\delta\Gamma_{1+2,l}&=&\left(\frac{\eta}{s}\right)\Bigg[\frac{A_{b,q}g^4}{(2\pi)^34\tau v_p}\left(
\frac{-v_p}{2}+\frac{v_p^3}{3}-\frac{v_p^5}{10}\right)\left(
\frac{\pi^2}{3}-\frac{3\zeta(3)}{2}\right)\left(-\frac{1}{2}+\ln\left|\frac{q_{max}}{m_D}\right|
\right)\nn\
&+&\frac{A_{b,q}g^4T^2}{(2\pi)^34\tau v_p}\left(\frac{-v_p^3}{3}+\frac{v_p^5}{5}\right)\left(\frac{7\pi^4}{15}-\frac{45\zeta(5)}{2}\right)
\left(\frac{1}{m_D^2}-\frac{1}{q^{2}_{max}}\right)\Bigg].
\eea


The magnetic interaction on the other hand gives rise to,
\bea
\delta\Gamma_{1+2,t}^{soft}&=&\left(\frac{\eta}{s}\right)\frac{\eta A_{b,q}g^4}{(2\pi)^38T^3\tau v_p}
\int_0^{\infty}k^4(f_k^0-2f_k^{02})dk\int_0^{q^*} 
dq\int_{-v_pq}^{v_pq}d\omega \left(\frac{1}{3}-\chi\right)\frac{\left(v_p^2-\frac{\omega^2}{q^2}\right)}{2\left(1-\frac{\omega^2}{q^2}\right)
\left(q^4+\frac{\pi ^2m^4\omega^2}{4q^2}\right)}
\nn\
&\simeq&\left(\frac{\eta}{s}\right)\frac{\eta A_{b,q}g^4T^2\pi}{(2\pi)^316\tau v_p m_D^2}
\left(\frac{7 \pi ^4}{15}-\frac{45 \zeta (5)}{2}\right)\int_0^{q^*} \frac{dq}{q}.
\eea

 The transverse part is infrared divergent even after using the HTL resummation and the form of divergence is same  
 as obtained in the non-viscous medium \cite{Blaizot96,Blaizot97}.

 Computation of the last two terms of the damping rate is presented below,
\bea
\delta\Gamma_{3+4,l}^{soft}&=&\left(\frac{\eta}{s}\right)\frac{\eta A_{b,q}g^4}{(2\pi)^32T^3\tau v_p}
\left(-\frac{ v_p^3}{3 }
+\frac{ v_p^5}{5 }
\right)
\int_0^{\infty}(2k^3f_k^0f_k'^0-k^2f_k^{02})dk
\int_0^{q^*}\frac{q^3dq}{(q^2+m_D^2)^2}\nn\
 &=&\left(\frac{\eta}{s}\right)\frac{ A_{b,q}g^4}{(2\pi)^32\tau v_p}\left(-\frac{v_p^3}{3 }+\frac{v_p^5}{5 }\right)
\left(\frac{\pi ^2}{3}  -3\zeta (3)\right)\left(-\frac{1}{2}+\ln\left|\frac{q^{*}}{m_D}\right|
\right)\nn\
\delta\Gamma_{3+4,l}&=&\left(\frac{\eta}{s}\right)\frac{ A_{b,q}g^4}{(2\pi)^32\tau v_p}
\left(-\frac{v_p^3}{3 }+\frac{v_p^5}{5 }\right)
\left(\frac{\pi ^2}{3} -3 \zeta (3)\right)\left(-\frac{1}{2}+\ln\left|\frac{q_{max}}{m_D}\right|
\right).
\eea
The transverse interaction on the other hand gives,
\bea
\delta\Gamma_{3+4,t}^{soft}=\left(\frac{\eta}{s}\right)\frac{ A_{b,q}g^4}{8(2\pi)^3\tau v_p}
\Bigg[\left(\frac{\pi ^2}{3}-3 \zeta (3)\right)
\left(-\frac{16v_p^5}{225}-\frac{4v_p^5}{15}\ln\left|\frac{2q_{max}}{m_D\sqrt{\pi v}}\right|\right)\Bigg].
\eea

Unlike the previous two terms here in this case we obtain finite interaction rate in the magnetic
 sector using the resummed propagator. This is because of the fact that one extra $\omega$
  in the numerator coming from the phase-space distribution function ($\omega \partial f_k/\partial k$) cures the logarithmic
   divergence. 
   
   One obtains same result both for the q-q and q-g scattering $t$ channel diagrams except the multiplicative group factor. $A_{b,g}$ differs from the quark case according to Eq.(\ref{mat_amp_bare_t}).

\subsection{Quark-gluon scattering}
In this section we present the detailed calculation of the contribution of quark-gluon scattering in the $s$ and $u$ channels 
to the total heavy quark damping rate. 
We start with the following expression, 
  \bea
   \delta \Gamma_{qg}^{\eta}&=&\left(\frac{\eta}{s}\right)\frac{8A_fg^4}{2T^3\tau}\int \frac{d^3kd^3k'd^3p'}{(2\pi)^52E_p2E_{p'}2k2k'}
    \left[\Phi_{k}f_k^0(1+f_{k'}^0)+
\Phi_{k'}f_{k^{'}}^0f_{k}^0\right]\nn\
&\times&\delta^4(P+K-P'-K')
   \left[\frac{-\tilde u}{\tilde s}+\frac{-\tilde s}{\tilde u}\right].
   \label{compt_vis_1}
   \eea

  In short notation we write the above equation as,
  \bea
    \delta \Gamma_{qg}^{\eta}&=&\left(\frac{\eta}{s}\right)\frac{8A_fg^4}{2T^3\tau}\int \frac{d^3kd^3k'd^3p'}{(2\pi)^52E_p2E_{p'}2k2k'}
    \left[\Phi_{k}f_k^0(1+f_{k'}^0)+
\Phi_{k'}f_{k^{'}}^0f_{k}^0\right]\nn\
&\times&\delta^4(P+K-P'-K')g(s,t,\omega),
    \label{compt_vis_1}
  \eea
  where, $g(s,t,\omega)$ depends on the Mandelstam variables and exchanged energy. The $k'$ integration can be 
expressed as
\bea
&&\int_{k'} \frac{\Phi_{k}f_k^0(1+f_{k'}^0)+
\Phi_{k'}f_{k^{'}}^0f_{k}^0}{2k'}\, (2\pi)^4 \delta^{(4)}(P+K-P'-K')\nn\
&=&2\pi \Phi_k f_k^0(1+f_{k-\omega}^0)+
\Phi_{k-\omega}f_{k-\omega}^0f_{k}^0 \Theta({k-\omega})\,
\delta((K-Q)^2) ,
\label{eq:k'-int}
\eea
  with the help of the following expression
  \[
	\int \frac{d^3 {\bm k}'}{(2\pi)^3}\, \frac1{2k'}
	=
	2\pi \int \frac{d^4k'}{(2\pi)^4}\, \Theta(k-\omega)\, \delta( (K-Q)^2) \, .
\]
To perform the integration over $p'$, $p$ is chosen along the $z$ axis and $k$ in the $y-z$ plane, hence,
\bea
	\bm p &=& (0,0,1)p \, , \nonumber \\
	\bm k &=& (0,\sin\theta_k,\cos\theta_k)k \, , \nonumber \\
	\bm p' &=& (\sin\theta_{p'}\sin\phi_{p'},\sin\theta_{p'}\cos\phi_{p'},\cos\theta_{p'})p' \, .
\label{eq:angles}
\eea
The integration over $\phi$ can be done with the help of the delta function as shown below,
\bea
\int_0^{2\pi} d\phi\, \delta((K-Q)^2) = \frac2{\sqrt{f}}\, \Theta(f) \, ,
\label{eq:phi-int}
\eea
where, $f=B^2-A^2$. $B$ and $A$ can be expressed in terms of the Mandelstam invariants \cite{Peigne08a,Peigne08b},
\bea
A &=& s -m_q^2 +t-2kE_{p'}+2kp'\cos\theta_k\cos\theta_{p'} \, , \nn  \
B &=& 2kp'\sin\theta_k\sin\theta_{p'} \, .
\label{AB}
\eea
  We now change the variables from $p'$ and $\cos\theta_{p'}$ to $t$ and $\omega$ respectively by the following transformation, 
\bea
t &=& 2(m_q^2 - E_pE_{p'} + p p' \cos\theta ), \nn\
\omega &=& E_p - E_{p'} \, .
\label{eq:change_variables}
\eea
With this Eq.(\ref{compt_vis_1}) now becomes,
\bea
\delta \Gamma_{qg}^{\eta}&=&\left(\frac{\eta}{s}\right)\frac{A_fg^4}{4T^3\tau \pi^2 pE_p}\int_k \frac{1}{2k}\int_{-\infty}^0 dt
\int_{-\infty}^{\infty} \frac{d\omega}{\sqrt{f(\omega)}}\,\nn\
&&\left(\Phi_{k}f_k^0(1+f_{k-\omega}^0)+
\Phi_{k-\omega}f_{k-\omega}^0f_{k}^0\right)g(s,t,\omega) \, .
\label{eq:I}
\eea
Bounds on the integrals $\omega$ and $t$ arise from the condition $f=B^2-A^2\geq 0$. $f(\omega)$ can now be written as follows \cite{Peigne08a,Peigne08b},
\bea
f(\omega) = -a^2\omega^2 + b\,\omega +c \, ,
\label{eq:g(omega)}
\eea
the coefficients of the above equation are \cite{Peigne08a,Peigne08b},
\bea
a &=& \frac{s-m_q^2}{p} \, , \nn  \
b &=& -\frac{2t}{p^2}( E_p(s-m_q^2) - k(s+m_q^2) ) ,\nn \
c &=& -\frac{t}{p^2} [ t( (E_p+k)^2-s ) + 4p^2k^2 -(s-m_q^2-2E_pk)^2 ] .
\label{abc}
\eea
$f(\omega)$ is positive only in the domain $\omega_{min}<<\omega<<\omega_{max}$, where the discriminant $D=4a^2c+b^2$ is positive. Thus
 we have \cite{Peigne08a,Peigne08b},
 \bea
&& \ \ \ \omega_{\rm min}^{\rm max} = \frac{b \pm \sqrt{D}}{2a^2} \, , \nn\
&& D = -t \left( st+(s-m_q^2)^2 \right) \left( \frac{4k\sin\theta_k}p \right)^2 \, .
 \eea
 The condition $D\geq 0$ leads to the $2\rightarrow 2$ scattering processes with one massless and one massive particle in the limit
  $t_{min}\leq t \leq 0$ with \cite{Peigne08a,Peigne08b},
  \bea
  t_{\rm min} = -\frac{(s-m_q^2)^2}{s}  \ \ .
  \eea
  We show latter in this section that the first term in Eq.(\ref{eq:I}) gives us the finite contribution, the other terms are
  either quadratic in $f_k$ or higher oreder in $\omega$. Hence, for the present purpose evaluation of the first term is sufficient.
 Considering only the first term we obtain,
 \bea
\delta \Gamma_{qg}^{\eta}&=&\left(\frac{\eta}{s}\right)\frac{A_fg^4}{4T^3\tau \pi^2 pE_p}\int_k \left(\frac{1}{3}-\cos^2\theta_{kz}\right)\frac{kf_k^0}{2}\int_{t_{min}}^0 dt
\int_{\omega_{min}}^{\omega_{max}} \frac{d\omega}{\sqrt{f(\omega)}} 
\left[\frac{-\tilde u}{\tilde s}+\frac{-\tilde s}{\tilde u}\right]\, .
\label{eq:II}
\eea
 Evaluation of the $\omega$ integral gives,
 \bea
I_\omega=\int_{\omega_{\rm min}}^{\omega_{\rm max}} d\omega\,\frac{1}{\sqrt{f(\omega)}}=
{\rm Re} \int_{-\infty}^\infty d\omega\, \frac{1}{\sqrt{f(\omega)}} \, =\frac{\pi}{a}.
\eea
Eq.(\ref{eq:II}) now becomes,
\bea
\delta \Gamma_{qg}^{\eta}&=&\left(\frac{\eta}{s}\right)\frac{A_fg^4}{4T^3\tau \pi E_p}\int_k \left(\frac{1}{3}-\cos^2\theta_{kz}\right)\frac{kf_k^0}{2}\int_{t_{min}}^0 dt
\frac{1}{s-m_q^2} \left[\frac{-\tilde u}{\tilde s}+\frac{-\tilde s}{\tilde u}\right]\, .
\label{eq:III}
\eea
Dominating logarithmic contribution from q-g scattering to the heavy quark damping rate comes from the domain 
$\tilde u_{min}<<\tilde u<< \tilde u_{max}$ giving rise to,
\bea
&&\int_{\tilde u_{min}}^{\tilde u_{max}}\frac{d\tilde u}{\tilde s} \left[\frac{-\tilde u}{\tilde s}+\frac{-\tilde s}{\tilde u}\right]\nn\
&&=\ln\left|\frac{s}{m_q^2}\right|+\frac{1}{2}-\frac{m_q^4}{2s^2}.
\eea
We now focus in the limit $E_p>>m_q^2/T$, which implies $s=m_q^2+2PK\sim {\cal O}(E_pT)>>m_q^2$. In this limit we can consider only the
 logarithmic term $\ln\left|\frac{s}{m_q^2}\right|$. The remaining $k$ integral can be evaluated as follows,
 \bea
&&\frac{1}{8\pi^2}\int_0^{\infty}k^3dkf_k^0\int_{-1}^1d\cos\theta_k\left(\frac{1}{3}-\cos^2\theta_{kz}\right)\ln\left|\frac{2E_pk(1-\cos\theta_{pk})+m_q^2}{m_q^2}\right|\nn\
 &=&\frac{1}{8\pi^2}\int_0^{\infty}k^3dkf_k^0\left(\frac{2}{9}-\frac{m_q^2}{3kE_p}\ln\left|1+\frac{4E_pk}{m_q^2}\right|\right)\nn\
&=&\frac{1}{8\pi^2}\left(\frac{2\pi^4T^4}{135}-\frac{T^3m_q^2}{3E_p}\left
(2\zeta(3)\ln\left|\frac{4E_pT}{m_q^2}\right|+3\zeta(3)-2\gamma\zeta(3)+2\zeta'(3)\right)\right).
 \eea
The final expression for the q-g scattering in the $s$ and the $u$ is now given by,
\bea
\delta \Gamma_{qg}^{\eta}&=&\left(\frac{\eta}{s}\right)\frac{A_fg^4}{32T^3\tau \pi^3 E_p}
\Bigg(\frac{2\pi^4T^4}{135}\nn\
&-&\frac{T^3m_q^2}{3E_p}\left
(2\zeta(3)\ln\left|\frac{4E_pT}{m_q^2}\right|+3\zeta(3)-2\gamma\zeta(3)+2\zeta'(3)\right).
\label{eq:IV}
\eea
\begin{acknowledgments}
Authors would like to thank S. Mallik for instructive discussion. 
S. Sarkar would like to thank P. Roy for critical reading of the manuscript. 
\end{acknowledgments}


\begin{thebibliography}{50}
\bibitem{Kovtun05} P. K. Kovtun, D. T. Son and A. O. Starinets 
Phys. Rev. Lett. {\bf 94}, 111601 (2005).
\medskip
\bibitem{Teaney03} D. Teaney, Phys. Rev. C {\bf 68}, 034913 (2003).
\bigskip
\bibitem{Dusling10} K. Dusling, G. D. Moore and D. Teaney, Phys. Rev. C {\bf 81}, 034907 (2010).
\bigskip

\bibitem{Soff01} S. Soff, S. A. Bass and Adrian Dumitru, Phys. Rev. Lett. {\bf 86}, 3981 (2001).
\bigskip
\bibitem{Adler01} C. Adler {\em et al.}, STAR Collaboration, Phys. Rev. Lett. {\bf 87}, 082301 (2001).
\bigskip
\bibitem{Acdox02} K. Acdox {\em et al.}, PHENIX Collaboration, Phys. Rev. Lett. {\bf 88}, 192302 (2002).
\bigskip

\bibitem{Dusling08} K. Dusling Nucl. Phys. A. {\bf 809}, 245 (2008).
\bigskip
\bibitem{Dusling10A} K. Dusling Nucl. Phys. A. {\bf 839}, 70 (2010).
\bigskip

\bibitem{Bhatt10} J. R. Bhatt, H. Mishra and V. Sreekanth, JHEP {\bf 11}, 106 (2010).
\bigskip
\bibitem{Jiang10} B. Jiang and J. Li, Nucl. Phys. A {\bf 847}, 268 (2010).
\bigskip
\bibitem{Das12} S. K. Das, V. Chandra and Jan-e Alam, arXiv {\bf 1210.3905v1} (2012).
\bigskip
\bibitem{Pisarski92}  R. D. Pisarski, Phys. Rev. D {\bf 46}, 1829 (1992).
\bigskip
\bibitem{Pisarski93} R. D. Pisarski, Phys. Rev. D {\bf 47}, 5589 (1993).
\bigskip
\bibitem{Blaizot96} J. P. Blaizot and E. Iancu, Phys. Rev. Lett {\bf 76}, 3080(1996).
\bigskip
\bibitem{Blaizot97}  J. P. Blaizot and E. Iancu, Phys. Rev. D {\bf 55}, 973(1997).
\bigskip
\bibitem{Peshier05}  A. Peshier, J. Phys. G {\bf 31}, 371 (2005).
\bigskip
\bibitem{Boyanovsky98} D. Boyanovsky, H. J. de Vega, R. Holman, S. P. Kumar and R. D. Pisarski, Phys. Rev. D {\bf 58}, 125009 (1998).
\bigskip
\bibitem{Manuel97} M.Le Bellac and C. Manuel, Phys. Rev. D {\bf 55}, 3215(1997).
\bigskip
\bibitem{Manuel00} C. Manuel, Phys. Rev. D {\bf 62}, 076009(2000).
\bigskip
\bibitem{Sarkar10} S. Sarkar and A. K. Dutt-Mazumder, Phys. Rev. D {\bf 82}, 056003(2010).
\bigskip
\bibitem{AKDM13} S. Sarkar and A. K. Dutt-Mazumder in preparation.
\bigskip
\bibitem{AKDM05} A. K. Dutt-Mazumder, Jan-e Alam, P. Roy and B. Sinha, Phys. Rev. D {\bf 71}, 094016 (2005).
\bigskip
\bibitem{Heiselberg93}  H. Heiselberg and C. J. Pethick, Phys. Rev. D. {\bf 48}, 2916(1993).
\bigskip
\bibitem{Heiselberg94}  H. Heiselberg, Phys. Rev. D. {\bf 49}, 4739(1994).
 \bigskip
\bibitem{Arnold00} P. Arnold, G. D. Moore and L. G. Yaffe JHEP {\bf 0011}, 001(2000).
\bigskip
\bibitem{Groot} S. de Groot, W. van Leevuen and Ch. van Veert, {\em Relativistic Kinetic Theory} (North-Holland, Amsterdem, 1980).
\bigskip
\bibitem{Arnold03} P. Arnold, G. D. Moore and L. G. Yaffe JHEP {\bf 0305}, 051(2003).
\bigskip
\bibitem{Pisarski89} R. D. Pisarski, Phys. Rev. Lett. {\bf 63}, 1129 (1989).
\bigskip
\bibitem{Pisarski90} E. Braaten and R. D. Pisarski, Phys. Rev. Lett. {\bf 64}, 1338(1990).
\bigskip
\bibitem{Braaten91} E. Braaten and R. D. Pisarski, Nucl. Phys. B {\bf 337}, 569(1970).
\bigskip
\bibitem{Yuan91} E. Braaten and T. C. Yuan, Phys. Rev. Lett. {\bf 66}, 2183(1991).
\bigskip
\bibitem{Beraudo06} A. Beraudo, A. De Pace, W.M. Alberico, A. Molinari, Nucl. Phys. A {\bf 831}, 59(2009).
 \bigskip
\bibitem{Thoma91} M. H. Thoma and M. Gyulassy, Nucl. Phys. B. {\bf 351}, 491(1991).
\bigskip
\bibitem{Peigne08a} S. Peigne and A. Peshier, Phys. Rev. D {\bf 77}, 014015 (2008).
\bigskip
\bibitem{Peigne08b} S. Peigne and A. Peshier, Phys. Rev. D {\bf 77}, 114017 (2008).


\end{thebibliography}
\end{document}